\documentclass[twocolumn,prl, showpacs]{revtex4-1}
\usepackage{graphicx}
\usepackage{dcolumn}
\usepackage{bm}
\usepackage[table]{xcolor} 

\begin{document}





\title{Ferrimagnetic ordering and spin entropy of\\ field-dependent intermediate spins in $Na_{0.82}CoO_2$}

\author{G. J. Shu$^1$}
\author{F. C. Chou$^{1,2,3}$}
 \email{fcchou@ntu.edu.tw}
\affiliation{
$^1$Center for Condensed Matter Sciences, National Taiwan University, Taipei 10617, Taiwan}
\affiliation{
$^2$National Synchrotron Radiation Research Center, Hsinchu 30076, Taiwan}
\affiliation{
$^3$Taiwan Consortium of Emergent Crystalline Materials, Ministry of Science and Technology, Taipei 10622, Taiwan}

\date{\today}

\begin{abstract}
The peculiar field-dependent magnetism of $Na_{0.82}CoO_2$ has been investigated through an analysis of its DC and AC spin susceptibilities.  To account for the easily activated narrow $b_{2g}$-$a_{1g}$ gap of the crystal field for Co in the cobalt oxide layer, the spin-state transition of Co$^{3+}$ (3d$^6$) between the low spin (LS) state $b_{2g}^2a_{1g}^0$ of S=0 and the intermediate spin (IS) state $b_{2g}^1a_{1g}^1$ of S=1 is thus seen as thermally activated and exhibits a Boltzmann distribution. The IS state of Co$^{3+}$ within each $\sqrt{13}a$ hexagonal superlattice formed by the S=1/2 state of the Co$^{4+}$ ions appears randomly within each supercell and shows significant temperature and field dependence.  The magnetic field is found to assist in pinning down the thermally activated state of Co$^{3+}$ and swings the Boltzmann distribution weight toward a higher fraction of the IS state.  
The field dependence of the in-plane magnetic moment from the added number of S=1 spins is used to explain the origin of A-type antiferromagnetic (AF) ordering, particularly that the ferromagnetic (FM)-like behavior below T$_N$ at low field is actually a ferrimagnetic IS spin ordering of Co$^{3+}$.

\end{abstract}



\maketitle


\begin{figure}
\includegraphics[width=3.5in]{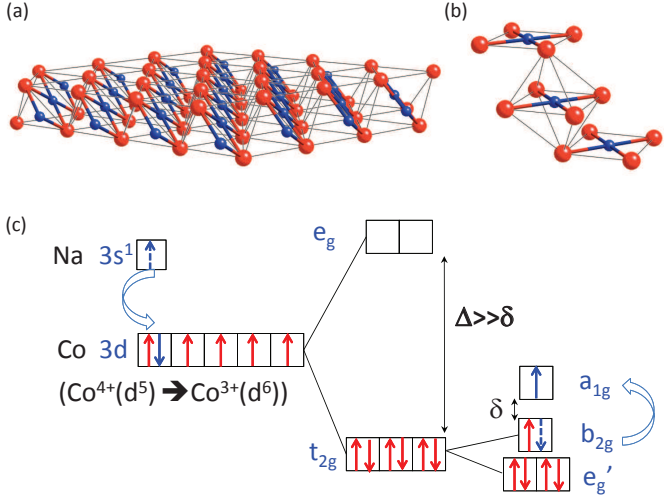}
\caption{\label{fig-hybridCEF}(color online) (a) Conventional view of the CoO$_2$ plane of Na$_x$CoO$_2$ as a layer of edge-sharing CoO$_6$ octahedra have been revised into weakly coupled edge-sharing CoO$_2$ chains. (b) The correct crystal field for Co has a square-planar shape with two apical oxygens that belong to the neighboring CoO$_2$ chains of weaker effective charge, see Ref.~\onlinecite{Shu2013} for details.  (c) The electronic configuration of Co$^{3+}$ with $d^6$ sitting in the proposed CEF has a narrow $b_{2g}$-$a_{1g}$ gap with an LS state of S=0 and an IS state of S=1. }
\vspace{-5mm}
\end{figure}

$Na_xCoO_2$ has been shown to have an intriguing and rich electronic phase diagram as a function of Na ions that are sandwiched between CoO$_2$ layers, including a superconducting state for x$\sim$ 1/3 after hydration,\cite{Takada2003} a metal-to-insulator transition for x$\sim$1/2, and a Curie-Weiss metal phase with A-type antiferromagnetic ordering below $T_N \sim 22K$ for $x$$\sim$0.82-0.86.\cite{Chou2008, Shu2009}  In addition, several large hexagonal superlattices were identified for $x$$\sim$0.71 ($\sqrt{12}$a) and 0.82-0.86 ($\sqrt{13}$a) due to Na vacancy/ion ordering, which has also be been shown to have a direct impact on the Fermi surface reconstruction.\cite{Balicas2008}  In addition, $Na_xCoO_2$ has also been introduced as a promising thermoelectric material,\cite{Terasaki1997} being used either at low-temperature for its large enhancement of thermopower near 50K for $x$ near 0.85,\cite{Lee2006} or at high temperature for its thermoelectric figure-of merit ($ZT$) higher than 1.\cite{Fujita2001}  The findings of large enhancement of thermopower and magnetic field suppression have been interpreted as being important evidence that spin entropy may play a key role in the Heikes form of the Seebeck coefficient (the thermopower),\cite{Wang2003} i.e., the spin entropy contribution to the Seebeck coefficient of a semiconductor can be suppressed by the magnetic field.  Alternative explanations such as strong electron correlations,\cite{Haerter2006}, combined disorder and electronic correlations,\cite{Wissgott2011}, and the narrow manifold of $t_{2g}$ bands in this system \cite{Xiang2007} have also been proposed.   

Conventionally, the magnetic properties of $Na_xCoO_2$ have often been discussed starting from the spins of mixed valence Co$^{3+}$/Co$^{4+}$ under the CoO$_6$ octahedral crystal field, where the low-spin (LS) state of d$^5$ for Co$^{4+}$ has S=1/2, and the LS state of d$^6$ for Co$^{3+}$ is nonmagnetic.\cite{Amatucci1996}  However,  \textit{ab initio} LDA calculations have yielded a reversed $t_{2g}$ splitting of $a_{1g}$-$e_g^\prime$ relative to the classical crystal electric field theory (CEF) prediction, in addition, the predicted $e_g^\prime$ hole pockets along $\Gamma-K$ at $k_z$ = 0 have never been observed by the ARPES experiments.\cite{Singh2000, Johannes2005, Landron2008, Yang2005}  We have resolved this issue by revising the view of CEF that the narrow $d$-orbital splitting should be assigned to be $b_{2g}$-$a_{1g}$ rigorously following the CEF theory, instead of the previously assumed $e_g^\prime$-$a_{1g}$, as summarized in Fig.~\ref{fig-hybridCEF}.\cite{Shu2013} It was proposed and verified that the small gap of $b_{2g}$-$a_{1g}$ $\sim$17 meV at H=1 Tesla could be subjected to easy thermal activation above $\sim$100 K, and the fraction of the thermally activated excited intermediate spin (IS) state of S=1 ($b_{2g}^1a_{1g}^1$) versus the ground state (LS) of S=0 ($b_{2g}^2a_{1g}^0$) for Co$^{3+}$ spins was shown to follow the Boltzmann distribution statistically. In this study, we explore the field dependence of the IS state of Co$^{3+}$ and use it to interpret the intriguing spin dynamics of $Na_{0.82}CoO_2$, including how and why the narrow $b_{2g}$-$a_{1g}$ gap size is tuned sensitively by the magnetic field, and why FM-like behavior is observed at low field but an A-type antiferromagnetic (AF) ordering is confirmed at high field.

\begin{figure}
\begin{center}
\includegraphics[width=3.5in]{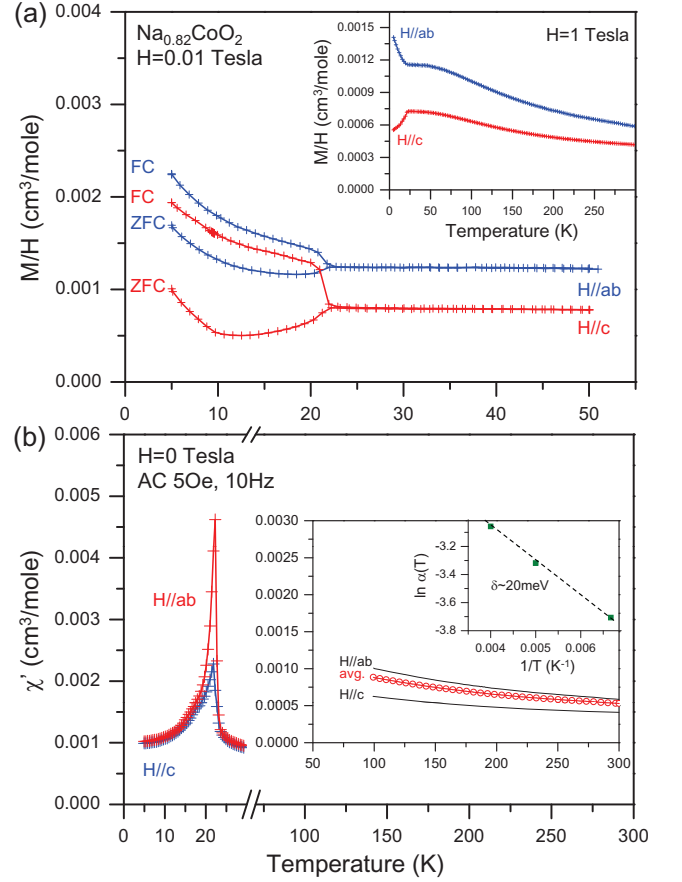}
\end{center}
\caption{\label{fig-chiT}(Color online) (a) Homogeneous spin susceptibilities ($\chi$=M/T) of single-crystal $Na_{0.82}CoO_2$ measured at a low applied field of 0.01 Tesla and a high field of 1 Tesla shown in the inset. While the high-field measurement shows the typical A-type antiferromagnetic transition of $T_N$$\sim$22K, the low-field measurement suggests a ferromagnetic-like transition with thermal and field hysteresis in both orientations. (b) AC spin susceptibilities measured using an AC field of 5 Oe/10Hz with zero applied field.  The insets show the powder-average ($\frac{2}{3}\chi^\|+\frac{1}{3}\chi^\bot$) data to extract $\alpha$(T), and the $\alpha$(T) in an Arrhenius plot to suggest a thermal activation gap of $\sim$20 meV.     }
\end{figure}

Based on the narrow $b_{2g}$-$a_{1g}$ gap from the CEF-lifted $3d$ degeneracy for Co$^{3+}$ (Fig.~\ref{fig-hybridCEF}(c)), the temperature-dependent Curie constant ($C=N \mu_{eff}^2/3k_B$) for $Na_{0.82}CoO_2$ has been revealed from the Curie-Weiss law ($\chi = \chi_\circ + \frac{C}{T-\Theta}$) analysis by using the homogeneous susceptibility $\chi$(T)=M/T data from different temperature sections above $\sim$5 T$_N$.\cite{Shu2013}  The Curie constants were found to be persistently higher than the value expected from the spin-only S=1/2 of Co$^{4+}$. In particular, the Curie constant could be interpreted to arise from three contributions, including the localized spins of Co$^{4+}$ in the LS state ($b_{2g}^1a_{1g}^0$) of S=1/2, Co$^{3+}$ in the LS state ($b_{2g}^2a_{1g}^0$) of S=0, and Co$^{3+}$ in the IS state ($b_{2g}^1a_{1g}^1$) of S=1.  The thermally activated IS state of Co$^{3+}$ for a system with a narrow $b_{2g}$-$a_{1g}$ gap of $\sim$17 meV (at H=1 Tesla) appears convincingly responsible for the intriguing magnetic properties.\cite{Shu2013}  

\begin{figure}
\begin{center}
\includegraphics[width=3.5in]{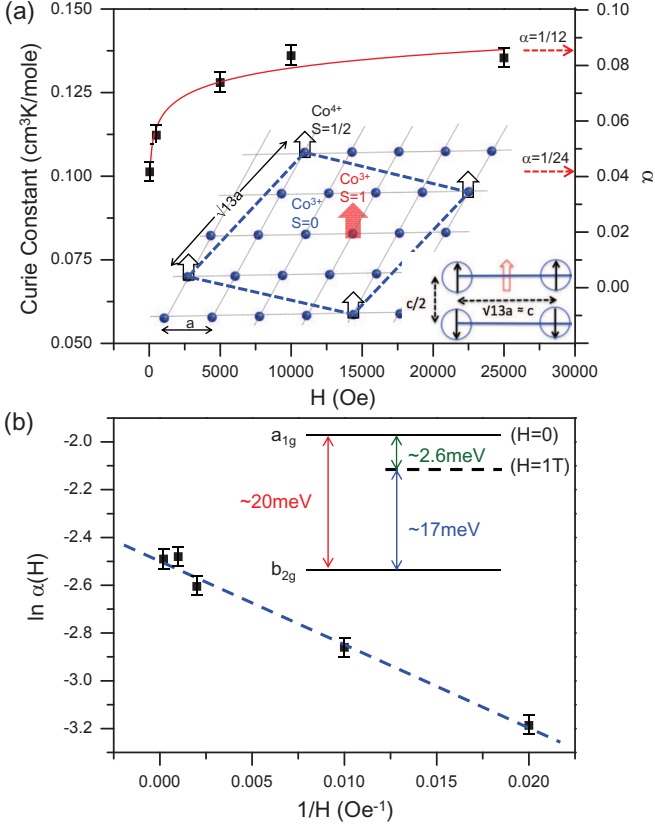}
\end{center}
\caption{\label{fig-ISactivation}(Color online)  (a) The Curie constant and the fraction of Co$^{3+}$ activated to the IS state at various fields, $\alpha$(H), increase with the field toward saturation. (b) Field-dependent activation behavior suggests a gap of $\sim$2.6 meV.  The inset of (a) shows top and side views of a superlattice cell of $\sqrt{13}$a, which correspond to the model of $\alpha$=1/12$\sim$0.083 and 1/24$\sim$0.042 for one IS per one and two superlattice cells, respectively. The inset of (b) shows a schematic diagram extracted from a measurement under 1 Tesla (see Ref.\onlinecite{Shu2013}) of the narrow $b_{2g}$-$a_{1g}$ gap in the Boltzmann distribution, where the gap size estimated from the zero field measurement $\sim$20 meV (see Fig.~\ref{fig-chiT}(b)) is reduced to $\sim$17 meV, as if the IS state level is shifted down by $\sim$2.6 meV effectively through the work done by the field.}
\end{figure}

Measurement results obtained by the same data analysis method previously applied to $Na_xCoO_2$,\cite{Shu2013} including the DC homogeneous susceptibilities ($\chi$=M/H) measured under various fields and a zero-field measurement using AC susceptibility measurement with H$^{AC}$=5 Oe/10 Hz, are shown in Fig.~\ref{fig-chiT}.  These high-field and low-field measurement results are consistent with those reported in the literature,\cite{Bayrakci2004, Luo2004} although the low-field FM-like behavior is drastically different from that of the A-AF ordering when measured at high field.\cite{Schulze2008, Shu2009}  
Using the common temperature range from 5T$_N$-300K that guarantees $k_B$T$>$$J_{nn}$ for a meaningful Curie-Weiss law fitting, we determined the Curie constant to be a superpositions of $C = C_{Co^{4+}}^{LS} + C_{Co^{3+}}^{LS} + C_{Co^{3+}}^{IS}$ from three types of localized spins, including the S=1/2 ($\mu_{eff}=1.732 \mu_B$) state of Co$^{4+}$, S=0 for Co$^{3+}$ in the ground LS state, and S=1 ($\mu_{eff}=2.828 \mu_B$) for Co$^{3+}$ in the excited IS state.  The fraction of Co$^{3+}$ activated to the IS state under various fields, $\alpha$(H), can thus be estimated from the relationship $\mu_{eff}^2 = (1-x)\times1.732^2+x(\alpha \times2.828^2+(1-\alpha)\times0^2)$ for $x$=0.82.  The Curie constant and the fraction of Co$^{3+}$ in the IS state ($\alpha$(H)) as a function of field are shown in Fig.~\ref{fig-ISactivation}.  The Curie constant grows with the applied field and approaches saturation at $\sim$7 Tesla.  When the extracted $\alpha$(H) is plotted against 1/H, as shown in the inset of Fig.~\ref{fig-ISactivation}, it is intriguing to note that an activation behavior with a gap of $\sim$2.6 meV is implied. It is clear that $\alpha$ is not only temperature-dependent, as noted previously,\cite{Shu2013} but displays field-dependence when the same temperature range is chosen in the paramagnetic regime. 

Consider the Co$^{3+}$ spin sitting in the energy spectrum, with a narrow $b_{2g}$-$a_{1g}$ gap; it is reasonable to assume that the magnetic field might pin a larger fraction of spins in the IS state. Because the $b_{2g}$-$a_{1g}$ gap size of $\sim$17 meV has previously been estimated under an applied field of H= 1 Tesla,\cite{Shu2013} the current field-induced activation gap of $\sim$2.6 meV implies that the actual gap size should be close to $\Delta$=17+2.6$\sim$20 meV when no field is applied.  We deduced the $\alpha$(H=0) value by using the AC susceptibility data shown in the inset of Fig.~\ref{fig-chiT}(b); the fitted gap size is $\sim$20 meV as expected. The field dependence of $b_{2g}$-$a_{1g}$ gap is summarized in Fig.~\ref{fig-ISactivation}, where the Boltzmann distribution can be interpreted as being revised by the applied field lowering the IS state by approximately 2.6 meV, mostly because of the Zeeman-like magnetic energy gain due to work done on the spin system by the applied field.    


The field dependence of the IS/LS fractional change for Co$^{3+}$ can also be interpreted quantitatively via the real space $\sqrt{13}a$ superlattice model for $Na_{0.82}CoO_2$.\cite{Chou2008, Shu2013} The $\sqrt{13}a$ hexagonal superlattice has been identified with the help of synchrotron X-ray Laue diffraction and electron diffraction as a result of the Na di-vacancy cluster ordering,\cite{Chou2008, Huang2009} and the Co atoms near the Na di-vacancy sites are proposed to be Co$^{4+}$ with S=1/2 to form a hexagonal superlattice of $\sqrt{13}$a size in the CoO$_2$ plane. Because the number and position of Co$^{4+}$ spins (S=1/2) are not expected to change with the applied field due to electron-phonon coupling via local distortion,\cite{Roger2007} as illustrated in Fig.~\ref{fig-ISactivation}(a), the observed field-dependent Curie constant must arise from the fractional change of the spin state for Co$^{3+}$ between the LS state of S=0 and the IS state of S=1.\cite{Shu2013}  It should be noted that the fraction of Co$^{3+}$ spins activated to the IS state for a specific temperature range and field, i.e., $\alpha$(T,H), has a constant magnitude but total freedom of distribution in space and time. For example, when one in twelve Co$^{3+}$ atoms per $\sqrt{13}$a superlattice cell is activated from the LS to the IS state, the position of the S=1 could be any Co$^{3+}$ at any instance, as if the Co$^{3+}$ atoms are LED bulbs that flash with a fixed total intensity.  It is clear that the thermal entropy from the randomness of the IS/LS state fluctuation for Co$^{3+}$ must introduce additional spin entropy to the system, in addition to the previously proposed configurational entropy between Co$^{3+}$/Co$^{4+}$.\cite{Wang2003} 

\begin{figure}
\begin{center}
\includegraphics[width=3.5in]{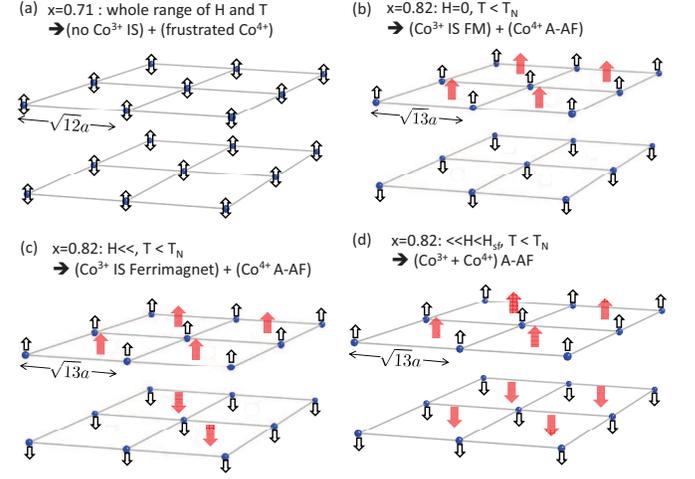}
\end{center}
\caption{\label{fig-HAAF}(Color online) (a) For $x$=0.71 of hexagonal superlattice size $\sqrt{12}$a without an IS state, where the blue arrows represent the localized spins of Co$^{4+}$ (S=1/2) at the corners, and all spins are geometrically frustrated.  (b)-(d) For $x$=0.82 on a hexagonal superlattice of size $\sqrt{13}$a with IS existence, the red arrows represent the excited IS spins of Co$^{3+}$ (S=1) appear randomly among the 12 Co$^{3+}$ ions per superlattice cell. (b) and (c) represent the ferrimagnetic ordering at zero and low field, respectively, and (d) illustrates the typical A-AF ordering at high field before the occurrence of spin-flop transition at H$_{sf}$. }
\end{figure}


The A-type AF spin ordering of $Na_xCoO_2$ (x $>$~0.71) with T$_N$$\sim$22K has been repeatedly confirmed by DC homogeneous susceptibility measurement under applied fields greater than $\sim$0.5 Tesla.\cite{Bayrakci2004, Luo2004}  However, FM-like behavior with different onsets near 8, 22, and 29 K has also been observed, with strong cooling rate and history dependence.\cite{Schulze2008, Alloul2012, Rhyee2008, Shu2009}  In addition, the staging effect for Na-layers with different stack orderings was found to be consistent with the highly sensitive  thermal and field history dependence for $x$ between $\sim$0.71-0.86.\cite{Shu2009}  We have examined the fitting of Weiss temperature ($\Theta$) from the PM-state spin susceptibility data obtained in fields from zero to 7 Tesla,\cite{Daghofer2006, Shu2013} and the $\Theta$ values were found to be persistently negative and nearly independent of the field.  As long as no triangular (or Kagome)-like geometric frustration mechanism occurs in the spin liquid-like systems,\cite{Balents2010} a negative $\Theta$$\sim$$\Sigma z_iJ_i$ fitted from the paramagnetic regime implies that an effective AF coupling must lead to a 3D long-range AF ordering at low temperature, and the observed FM-like behavior observed at low field seems to contradict the negative $\Theta$ value.\cite{White}  

To examine the puzzling field-dependent spin ordering, we have repeated the DC susceptibility measurements under high and low fields for $Na_{0.82}CoO_2$, including a zero-field AC (5 Oe/10 Hz) susceptibility measurement, as shown in Fig.~\ref{fig-chiT}.  A spin state transition near $\sim$22 K is found consistently; however, the typical A-AF behavior appears only at high field, but the low field measurement indicates FM-like behavior at the same onset with significant ZFC/FC hysteresis. Both the magnetization M(T) with thermal hysteresis at low field and the $\chi^\prime$(T$_N$) cusp of AC susceptibility at zero field (Fig.~\ref{fig-chiT}(b)) suggest that the order parameter is magnetization, as expected for a FM-like transition. In contrast, the negative Weiss temperature $\Theta$ fitted from the high temperature paramagnetic regime implies that the neighboring spins must  be antiferromagnetically aligned in the ground state, which leads to the possibility of a 3D long range ferrimagnetic ordering, i.e., coupling in which neighboring spins are AF coupled with incomplete cancellation due to different spin sizes.\cite{White}   

Considering the intra-plane inter-Co$^{4+}$ spin distance for $x$=0.82 ($\sqrt{13}$a $\sim$ c) and $x$=0.71 ($\sqrt{12}$a $\sim$ c) of triangular symmetry, and considering the negative $\Theta$ to imply stronger AF coupling for spins in the neighboring planes of inter-plane distance $c/2$, geometric spin frustration is expected for the localized Co$^{4+}$ S=1/2 spins, as illustrated in Fig.~\ref{fig-HAAF}(a) for $x$=0.71. Indeed, nearly perfect spin frustration has been observed for $x$=0.71, i.e., $x$=0.71 does not show any magnetic ordering down to $\sim$1 K, as if spin frustration persists like a spin liquid.\cite{Balents2010, Chou2008}  In contrast, $x$=0.82 shows FM-like behavior at low field and A-AF ordering at high field.  In addition to the insignificant superlattice size difference, the main difference between $x$=0.82 and $x$=0.71 is the percentage of IS spin activation for Co$^{3+}$, i.e., $\alpha$ has been found to be zero for $x$=0.71 over the entire temperature and field range,\cite{Shu2013} but it is non-zero and field-dependent for $x$=0.82, as shown in Ref.\onlinecite{Shu2013} and Fig.~\ref{fig-ISactivation}.  

Here, we propose an interpretation of the field-dependent magnetic ordering behavior of $Na_{0.82}CoO_2$, both qualitatively and quantitatively, that especially considers why A-AF ordering occurs at high field, but ordering is FM-like at low field. There are 12 Co$^{3+}$ per $\sqrt{13}$a superlattice cell (see Fig.~\ref{fig-ISactivation}(a)), and $\alpha$=$\frac{1}{12}$$\cong$8.3$\%$ corresponds to one in twelve Co$^{3+}$ per superlattice cell in the IS state, which naturally leads to a picture in which $\alpha$(H=0)$\cong$4.5$\%$ represents one IS in every two $\sqrt{13}$a superlattice cells, as illustrated in Fig.~\ref{fig-HAAF}(b).  In particular, the S=1 spins (IS) of Co$^{3+}$ exist in every other layer and couple ferromagnetically to the S=1/2 of Co$^{4+}$ with an A-AF ordering below T$_N$, which could be viewed as a superposition of a 3D FM ordering of IS for Co$^{3+}$ and a 3D A-AF ordering for Co$^{4+}$ of S=1/2.  The existence of IS for Co$^{3+}$ is clearly responsible for the observed FM-like behavior due to the ferromagnetic coupling between the S=1 of Co$^{3+}$ and the S=1/2 of Co$^{4+}$, which has been described as a Hund's rule coupling by Daghofer \textit{et al.} based on a spin-orbital-polaron model argument.\cite{Daghofer2006}  Such zero field FM behavior has been confirmed using AC susceptibilities, as shown in Fig.~\ref{fig-chiT}(b), where $\chi^\prime$ shows a sharp cusp near T$_N$ only due to the large FM magnetization that cannot be flipped in-phase with the small AC field.  These results are consistent with the A-AF spin structure proposed and with the zero-field neutron diffraction experiment via spin wave modulation,\cite{Bayrakci2005} where the FM contribution from the dilute S=1 (IS) of Co$^{3+}$ in every other layer cannot be resolved from the A-type AF ordering of Co$^{4+}$ along the c-direction anyway.      

To understand the spin dynamics of $x$=0.82 better, it is easier to separate the spins into two classes, the localized S=1/2 of Co$^{4+}$ at the corner of each $\sqrt{13}$a superlattice  and the spins of Co$^{3+}$ under IS/LS fluctuation.  The increasing field would introduce additional IS into the layers that initially lacked IS, and as long as no spin flop occurreds for H $<$ H$_{sf}$,\cite{Bayrakci2005}, an evolution from the lowest population of $\alpha$(H=0)=$\frac{1}{24}$$\approx$4.2$\%$ (i.e., one IS in every other superlattice cell) to the saturated level of $\alpha$(H$\gg$)=$\frac{1}{12}$$\approx$8.3$\%$ (i.e., one IS per superlattice cell) would be expected.  The two extreme cases are shown in Fig.~\ref{fig-HAAF}(b) and (d) as a mixed system of (FM IS for Co$^{3+}$)+(A-AF for Co$^{4+}$) for the former and a perfect A-AF for the latter.  It is interesting to note that when $\frac{1}{24}$$\lesssim$$\alpha$(H)$\lesssim$$\frac{1}{12}$ in the low field condition, the incomplete AF cancellation of the antiferromagnetically aligned IS between neighboring layers would lead to a \underline{ferrimagnetic ordering of IS}, as illustrated in Fig.~\ref{fig-HAAF}(c).  Considering the random nature of the IS state of Co$^{3+}$ in space and time, the in-plane FM coupling among Co$^{4+}$ spins could also be viewed as an effective Kondo coupling mediated by the itinerant electrons,\cite{Chou2008} although no conventional itinerant electrons are involved here; rather, randomly selected Co$^{3+}$ electron sites fluctuate between the levels of a narrow gap ($b_{2g}^2a_{1g}^0$$-$$b_{2g}^1a_{1g}^1$) as a result of $t_{2g}$ degeneracies lifted by the crystal field.  

           

In summary, following the modified square-planar crystal field description of Co in the $CoO_2$ plane of $Na_{0.82}CoO_2$, the narrow $b_{2g}$-$a_{1g}$ gap for Co$^{3+}$ allows spin excitation from the LS (S=0) to the IS (S=1) state with sensitive thermal activation under the influence of field.  The applied magnetic field was found to be effective in pinning an increased population in the IS state and shifting the weight of Boltzmann distribution. 
The ferromagnetically coupled spins of field-dependent IS for Co$^{3+}$ and the S=1/2 for Co$^{4+}$ order at low temperature into an A-type AF ordering at high field but a ferrimagnetic ordering at low field. In particular, depending on the magnetic field, the existence of IS is a necessary condition to induce the required in-plane FM spin coupling among the localized spins of S=1/2 for Co$^{4+}$ for the 3D ferrimagnetic or A-AF ordering below T$_N$.  Since the A-type AF ordering is proposed and confirmed to exist under high field, but early neutron scattering experiments were performed in zero field without considering the existence of IS, we believe a comparative neutron scattering with and without applied field should help to verify the IS model proposed in this study. 

We thank P. A. Lee for valuable comments on the Curie constant analysis.  FCC acknowledges the support from Ministry of Science and Technology in Taiwan under project number MOST-102-2119-M-002-004 and Academia Sinica under project number AS-100-TP2-A01.   GJS acknowledges the support provided by MOST-Taiwan under project number 103-2811-M-002 -001.


\begin{thebibliography}{99} 
\bibitem{Takada2003} K. Takada, H. Sakurai, E. Takayama-Muromachi, F. Izumi, R. A. Dilanian, and T. Sasaki, Nature \textbf{422}, 53 (2003).  
\bibitem{Chou2008} F. C. Chou, M.-W. Chu, G. J. Shu, F.-T. Huang, W. W. Pai, H.-S. Sheu, and P. A. Lee, Phys. Rev. Lett. \textbf{101}, 127404 (2008).
\bibitem{Shu2009} G. J. Shu, F.-T. Huang, M.-W. Chu, J.-Y. Lin, P. A. Lee, and F. C. Chou, Phys. Rev. B \textbf{80}, 014117 (2009).
\bibitem{Balicas2008} L. Balicas, Y.-J. Jo, G. J. Shu, F. C. Chou, and P. A. Lee, Phys. Rev. Lett. \textbf{100}, 126405 (2008).
\bibitem{Terasaki1997} I. Terasaki, Y. Sasago, and K. Uchinokura, Phys. Rev. B \textbf{56}, 12685(R) (1997).
\bibitem{Lee2006} M. Lee, L. Viciu, L. Li, Y. Wang, M. L. Foo, S. Watauchi, R. A. Pascal Jr., R. J. Cava, and N. P. Ong, Nature Materials \textbf{5}, 537-540 (2006).
\bibitem{Fujita2001} K. Fujita, T. Mochida, and K. Nakamura, Jpn. J. Appl. Phys. \textbf{40}, 4644��647 (2001).
\bibitem{Wang2003} Y. Wang, N. S. Rogado, R. J. Cava, and N. P. Ong, Nature \textbf{423}, 425��28 (2003).
\bibitem{Haerter2006}J. O. Haerter, M. R. Peterson, and B. S. Shastry, Phys. Rev. Lett. \textbf{97}, 226402 (2006).
\bibitem{Wissgott2011} P. Wissgott, A. Toschi, G. Sangiovanni, and K. Held, Phys. Rev. B \textbf{84}, 085129 (2011).
\bibitem{Xiang2007} H. J. Xiang and D. J. Singh, Phys. Rev. B \textbf{76}, 195111 (2007).   
\bibitem{Amatucci1996} G.G. Amatucci, J.M. Tarascon, and L.C. Klein, Journal of the Electrochemical Society \textbf{143}, 1114 (1996).
\bibitem{Singh2000} D. J. Singh, Phys. Rev. B \textbf{61}, 13397 (2000).
\bibitem{Johannes2005} M. D. Johannes, I. I. Mazin, and D. J. Singh, Phys. Rev. B \textbf{71}, 214410 (2005).
\bibitem{Landron2008} S. Landron and M.-B. Lepetit, Phys. Rev. B \textbf{77}, 125106 (2008).
\bibitem{Yang2005} H.-B. Yang, Z.-H. Pan, A. K. P. Sekharan, T. Sato, S. Souma, T. Takahashi, R. Jin, B. C. Sales, D. Mandrus, A. V. Fedorov,  Z. Wang, and H. Ding, Phys. Rev. Lett. \textbf{95}, 146401 (2005).
\bibitem{Shu2013} G. J. Shu and F. C. Chou, Phys. Rev. B \textbf{88}, 155130 (2013).
\bibitem{Bayrakci2004} S. P. Bayrakci, C. Bernhard, D. P. Chen, B. Keimer, R. K. Kremer, P. Lemmens, C. T. Lin, C. Niedermayer, and J. Strempfer, Phys. Rev. B \textbf{69}, 100410(R) (2004).
\bibitem{Luo2004} J. L. Luo, N. L. Wang, G. T. Liu, D. Wu, X. N. Jing, F. Hu, and T. Xiang, Phys. Rev. Lett. \textbf{93},187203 (2004). 
\bibitem{Schulze2008} T. F. Schulze, P. S. Hafliger, Ch. Niedermayer, K. Mattenberger, S. Bubenhofer, and B. Batlogg, Phys. Rev. Lett. \textbf{100}, 026407 (2008).
\bibitem{Huang2009} F.-T. Huang, M.-W. Chu, G. J. Shu, H. S. Sheu, C. H. Chen, L.-K. Liu, P. A. Lee, and F. C. Chou, Phys. Rev. B \textbf{79}, 014413 (2009).
\bibitem{Roger2007} M. Roger, D. J. P. Morris, D. A. Tennant, M. J. Gutmann, J. P. Goff, J. U. Hoffmann, R. Feyerherm, E. Dudzik, D. Prabhakaran, and A. T. Boothroyd, Nature \textbf{445}, 631 (2007).
\bibitem{Alloul2012} H. Alloul, I. R. Mukhamedshin, A. V. Dooglav, Ya. V. Dmitriev, V.-C. Ciomaga, L. Pinsard-Gaudart, and G. Collin, Phys. Rev. B \textbf{85}, 134433 (2012) .
\bibitem{Rhyee2008} J.-S. Rhyee, J. B. Peng, C. T. Lin, and S. M. Lee, Phys. Rev. B \textbf{77}, 205108 (2008).
\bibitem{Daghofer2006} M. Daghofer, P. Horsch, and G. Khaliullin, Phys. Rev. Lett. \textbf{96}, 216404 (2006).
\bibitem{Balents2010} L. Balents, Nature \textbf{464}, 199 (2010).
\bibitem{White} R. M White, Quantum Theory of Magnetism, 3rd Ed., Springer. 
\bibitem{Bayrakci2005} S. P. Bayrakci, I. Mirebeau, P. Bourges, Y. Sidis, M. Enderle, J. Mesot, D. P. Chen, C. T. Lin, and B. Keimer, Phys. Rev. Lett. \textbf{94}, 157205 (2005).



\end{thebibliography}

\end{document}